\documentclass[traditabstract]{aa} 
\usepackage{graphicx}
\usepackage{txfonts}
\usepackage{natbib}
\bibpunct{(}{)}{;}{a}{}{,}
\begin{document}
\def\lsun{L_{\sun}}
\def\msun{M_{\sun}}
\def\kms{\mathrm{km\,s}^{-1}}

   \title{Fragmentation and Kinematics of dense molecular cores in the filamentary infrared-dark cloud G011.11$-$0.12\thanks{Based on observations carried out with the IRAM Plateau de Bure Interferometer and the IRAM 30m Telescope. IRAM is supported by INSU/CNRS (France), MPG (Germany) and IGN (Spain).}}

   \author{Sarah E. Ragan
          \inst{\ref{mpia}}
          \and
          Thomas Henning
          \inst{\ref{mpia}}
	      \and
	      Henrik Beuther
     	  \inst{\ref{mpia}}
	      \and
	      Hendrik Linz
	      \inst{\ref{mpia}}
	      \and
          Sarolta Zahorecz
          \inst{\ref{mpia},\ref{eso}}
          }

   \institute{Max Planck Institute for Astronomy,
              K\"onigstuhl 17, 69117 Heidelberg, Germany\\
              \email{ragan@mpia.de} \label{mpia}
		  \and
		  European Southern Observatory,
		  Karl-Schwarzschild-Str. 2, 85748 Garching bei M\"unchen, Germany \label{eso}
             }

   \date{Received ; accepted }

\authorrunning{Ragan et al.}
\titlerunning{IRAM observations IRDC cores}

\abstract{We present new Plateau de Bure Interferometer observations of a region in the filamentary infrared-dark cloud (IRDC) G011.11-0.12 containing young, star-forming cores. In addition to the 3.2\,mm continuum emission from cold dust, we map this region in the N$_2$H$^+$(1-0) line to trace the core kinematics with an angular resolution of 2$''$ and velocity resolution of 0.2\,km~s$^{-1}$. These data are presented in concert with recent {\em Herschel} results, single-dish N$_2$H$^+$(1-0) data, SABOCA 350\,$\mu$m continuum data, and maps of the C$^{18}$O (2-1) transition obtained with the IRAM 30\,m telescope.  We recover the star-forming cores at 3.2\,mm continuum, while in N$_2$H$^+$ they appear at the peaks of extended structures. The mean projected spacing between N$_2$H$^+$ emission peaks is 0.18\,pc, consistent with simple isothermal Jeans fragmentation. The 0.1\,pc-sized cores have low virial parameters on the criticality borderline, while on the scale of the whole region, we infer that it is undergoing large-scale collapse. The N$_2$H$^+$ linewidth increases with evolutionary stage, while CO isotopologues show no linewidth variation with core evolution. Centroid velocities of all tracers are in excellent agreement, except in the starless region where two N$_2$H$^+$ velocity components are detected, one of which has no counterpart in C$^{18}$O. We suggest that gas along this line of sight may be falling into the quiescent core, giving rise to the second velocity component, possibly connected to the global collapse of the region.}
  
 \keywords{Stars: formation - ISM: clouds -
ISM: kinematics and dynamics
               }
\maketitle

%

\section{Introduction}

If we are to understand the enrichment and evolution of galaxies and their interstellar medium (ISM), the conditions under which high-mass stars form is an essential component of any theory to get right. They dominate the stellar energy and momentum feedback, which in turn drives the chemistry and thermodynamic state of the ISM, and thus determine the properties of further generations of star formation. 

Any theoretical model of high-mass star and cluster formation profoundly depends on the imposed initial conditions, thus it is of principle importance that observers diligently characterise the physical conditions of the earliest phases of the process on both large and small scales. All stars form in molecular clouds, confined to the densest regions within the clouds. In order for massive stars to form, a large amount of mass at relatively high concentration is required. The class of objects known as infrared-dark clouds (IRDCs), which appear in silhouette against the mid-infrared background of the Galaxy \citep{Perault1996,carey_msx}, have been identified as excellent hunting grounds for the precursors to high-mass star and cluster formation. 

The target of our study is the filamentary IRDC G011.11$-$0.12 (henceforth G11). Based on the \citet{Reid2009} Galactic rotation curve model, we find a kinematic distance of 3.41\,kpc to G11 (assuming a $v_\mathrm{lsr}$ = 29.2\,km\,s$^{-1}$) and adopt it throughout this work. We note that this distance differs from the distance inferred from extinction studies which infer distances of 4.1\,kpc \citep{Kainulainen2011} and 4.7\,kpc \citep{Marshall2009}. 	G11 was first found because of its remarkable absorbing contrast at 8\,$\mu$m (\cite{egan_msx}; see also \cite{Kainulainen2013}) and carries on in absorption up to 100\,$\mu$m \citep{A&ASpecialIssue-Henning}. This massive ($M (N > 10^{22}\,cm^{-2}$) $\sim$ 20000\,$\msun$) filament (projected length $\sim$ 30\,pc) exceeds the critical mass-per-unit length value for an isothermal cylinder \citep{Ostriker1964}, though, as with most IRDCs, G11 appears to be dominated by non-thermal motions \citep{Kainulainen2013}. {\em Herschel} observations have also revealed a population of embedded protostellar cores following the dense centre of the filament. The continuum peak `P1' \citep{Carey2000, Johnstone_G11} is a site of active high-mass star formation \citep{Pillai_G11, Gomez2011, Wang2014}.

\begin{figure*}
\includegraphics[width=2.7in]{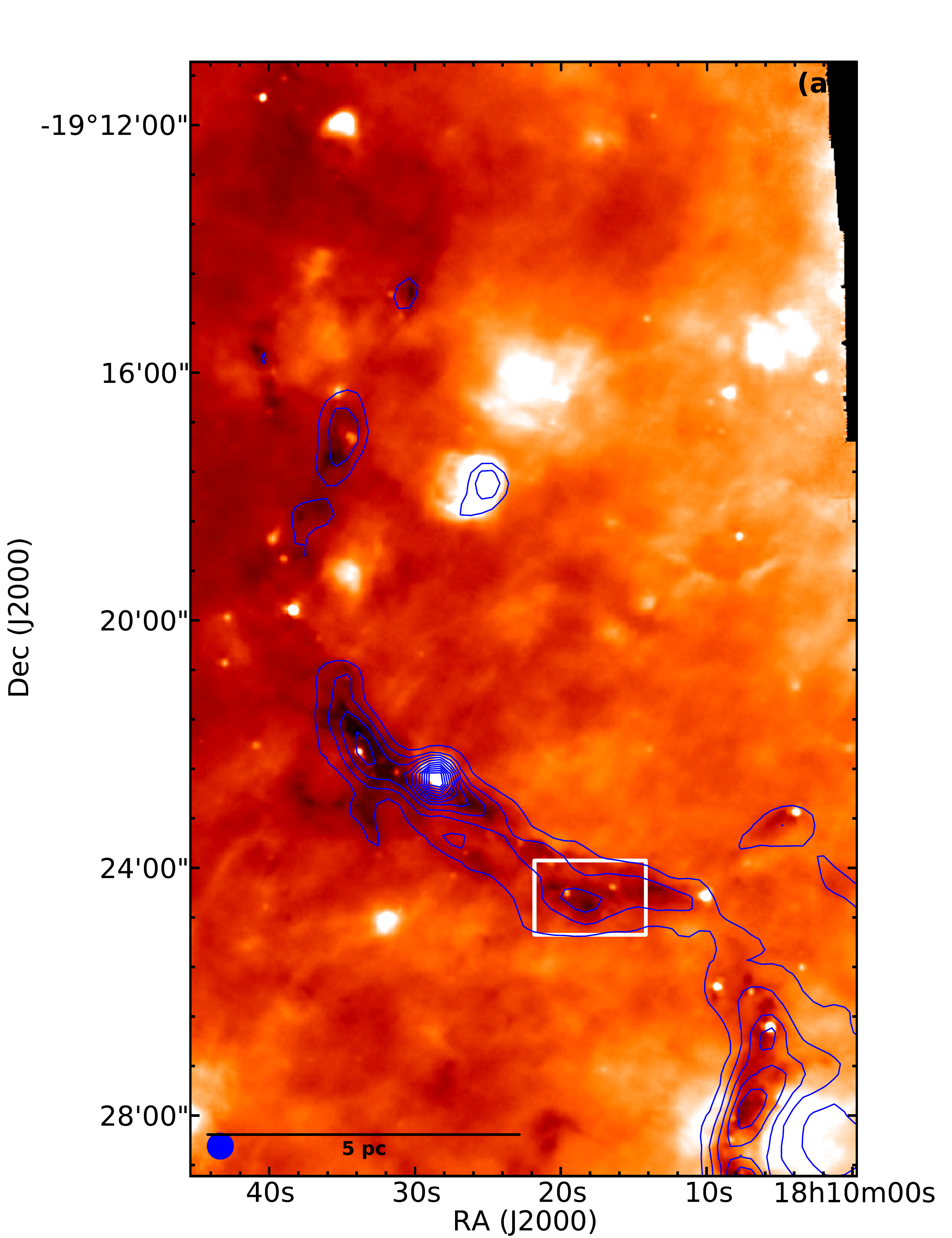}
\hspace{-0.32in}
\includegraphics[width=2.7in]{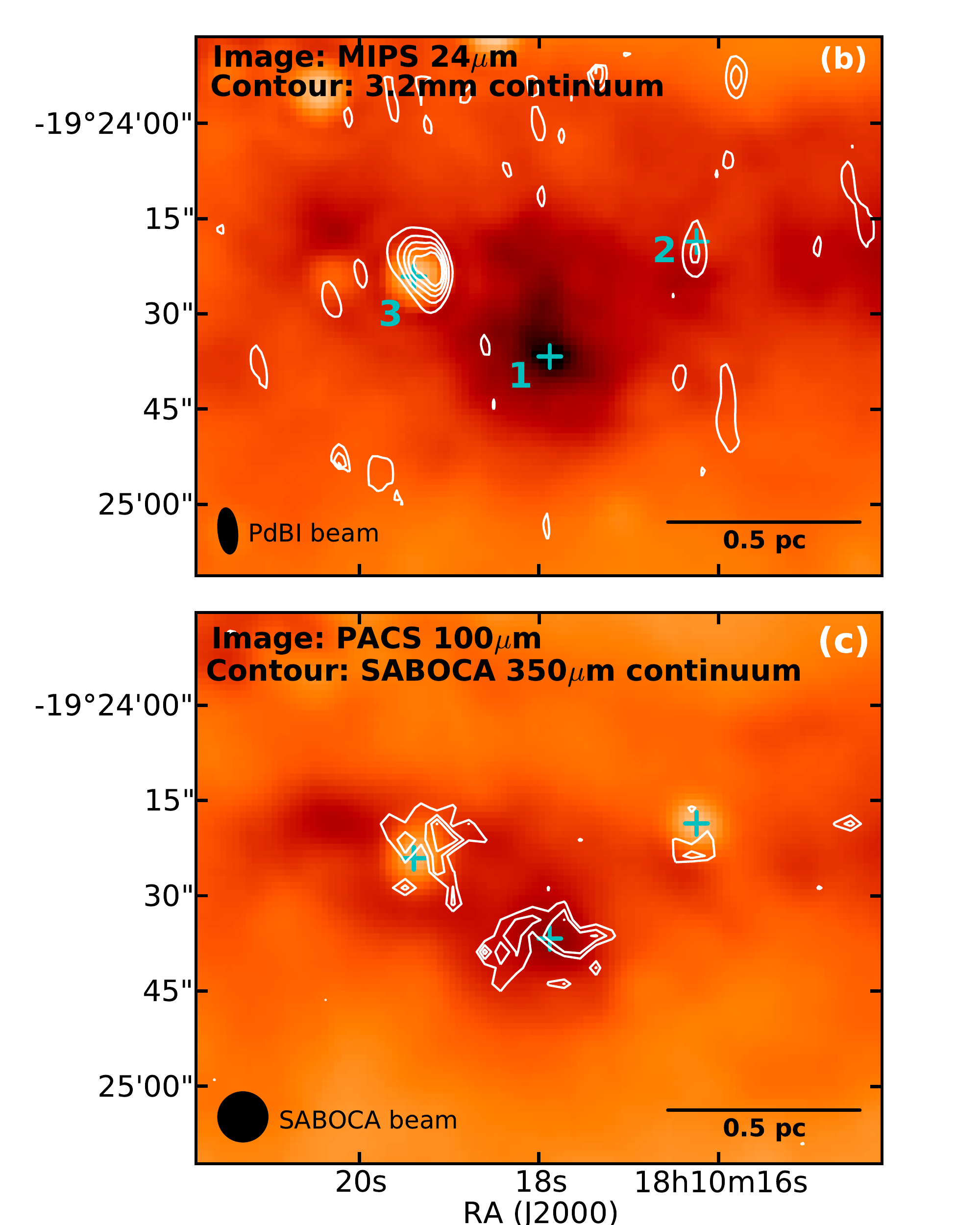}
\hspace{-0.32in}
\includegraphics[width=2.7in]{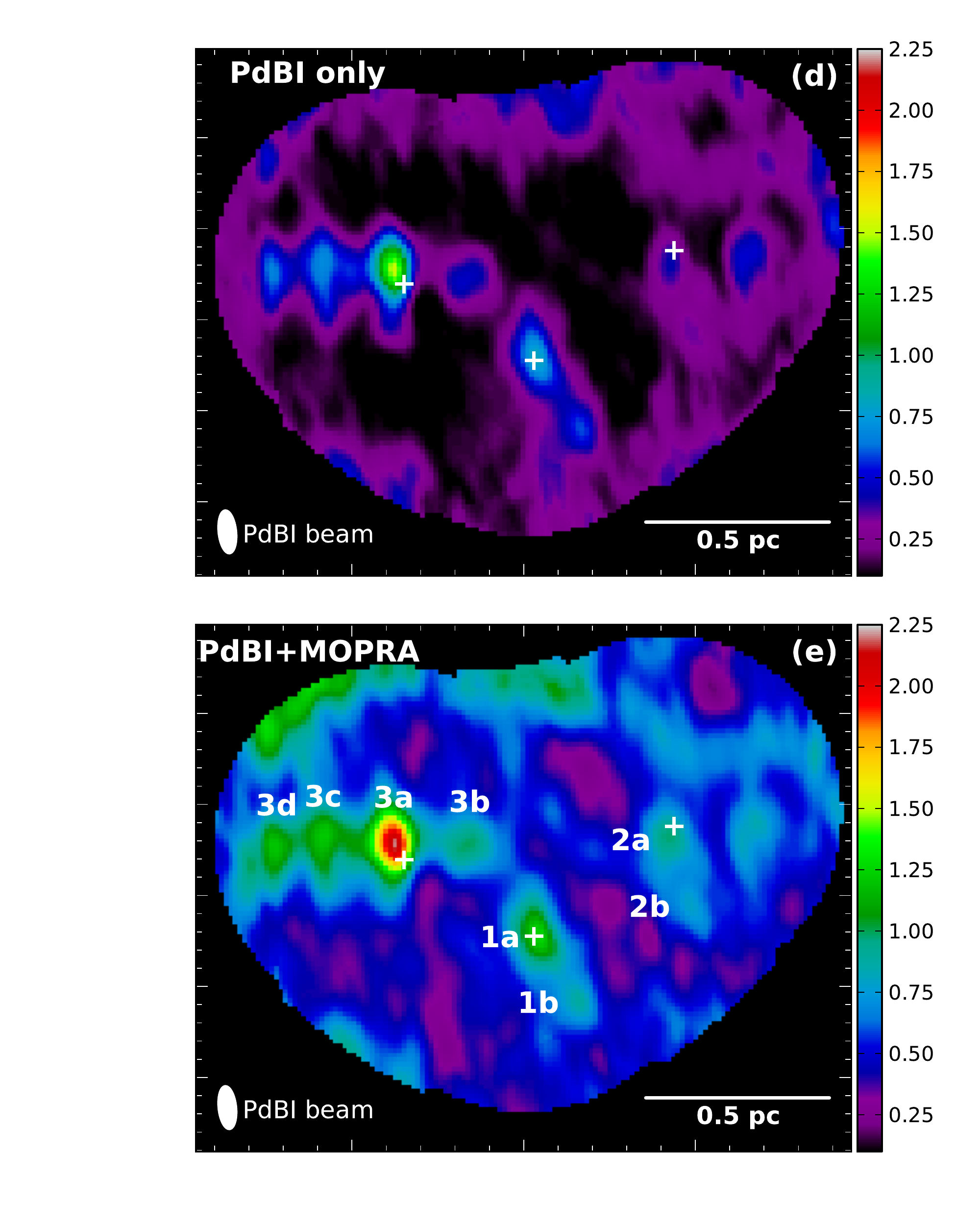}
\caption{\label{fig:finder}
\textbf{(a)}: The full G11 filament seen at 70\,$\mu$m with {\em Herschel}/PACS with SPIRE 350\,$\mu$m contours over-plotted from 9\,Jy beam$^{-1}$ increasing in steps of 3\,Jy beam$^{-1}$ \citep[adapted from ][]{A&ASpecialIssue-Henning}. The SPIRE 350\,$\mu$m beam (FWHM) is shown in the lower-left corner. The white rectangle marks the area shown in the rest of the panels.  \textbf{(b)} MIPS 24\,$\mu$m image with  3.2mm continuum contours from a PdBI-only map. The contour levels are from 0.2 to 0.12\,mJy\,beam$^{-1}$ in steps of 0.2\,mJy\,beam$^{-1}$. The synthesised beam is shown in the lower-left corner. \textbf{(c)} APEX-SABOCA 350\,$\mu$m continuum contours \citep{Ragan2013} at 0.81, 1.08, 1.35\,Jy\,beam$^{-1}$ plotted over the PACS 100\,$\mu$m image. The SABOCA 350\,$\mu$m beam (FWHM) is shown in the lower left. \textbf{(d)}  Integrated intensity of N$_2$H$^+$ (1-0) from the PdBI-only map. \textbf{(e)} Integrated intensity of the combined PdBI+MOPRA map. Both (d) and (e) are integrated from 29 to 33\,$\kms$, and the units of the colourscale are Jy beam$^{-1}$\,$\kms$. The $+$ signs indicate the positions of the main continuum cores identified in panel (b) and Table~\ref{tab:her_sab}.}
\end{figure*}

Figure~\ref{fig:finder}a shows the {\em Herschel} 70\,$\mu$m image first presented in \cite{A&ASpecialIssue-Henning} with the region we selected for this study shown in a box.  We chose this small region because it (a) appears relatively isolated from the feedback effects from the active massive star formation and (b) contained embedded cores of different inferred evolutionary stages \citep{Ragan2012b}. We present new observations with the Plateau de Bure Interferometer (PdBI) of both the 3.2\,mm continuum and N$_2$H$^+$(1-0) hyperfine transition. In addition, a new IRAM 30\,m map of the C$^{18}$O\,2$\rightarrow$1 transition is presented as a probe of the more diffuse environment. The goal of this study is to explore how G11 has fragmented into core substructures and whether there are distinct dynamical signatures -- in the dense or diffuse gas -- tied to the core evolution.

The new observations and those drawn from the literature are described in more detail in Section~\ref{s:obs}. We discuss the morphological and kinematic properties of the dense cores in Section~\ref{s:dense} and compare them to the environment probed by  C$^{18}$O in Section~\ref{s:environs}. We place our results in context with other studies of IRDCs in Section~\ref{s:disc} and summarise in Section~\ref{s:conclusion}. 

\section{Observations}
\label{s:obs}

\subsection{IRAM Plateau de Bure: N$_2$H$^+$\,1$\rightarrow$0}
\label{ss:n2hp}

We observed a subregion of G11 (see Figure~\ref{fig:finder}) in a three-point mosaic with the Plateau de Bure Interferometer (PdBI) in the D (June 2010) and C configuration (November 2010), which covered projected baselines from 13 to 180\,m as part of project U00D.  We tuned to the 93.174\,GHz transition of N$_2$H$^+$\,(1-0) in the lower sideband and, simultaneously, the $^{13}$CS\,(2-1) transition at 92.494\,GHz, which was not detected. Our spectral resolution is 0.2\,km s$^{-1}$. The maximum detectable source size is defined by the shortest baseline, $\lambda$/$D_\mathrm{min}$ = 28$''$, or 0.5\,pc in G11. 
 
The PdBI observations were reduced with the standard methods using the CLIC and MAPPING modules in the GILDAS software package \footnote{\texttt{http://www.iram.fr/IRAMFR/GILDAS}}. We corrected for phase and amplitude fluctuations with frequent observations of J1830-210, J1832-206, and 3C273, and the amplitude scale was set by observations of MWC349 ($S_{\nu}$ = 1.15\,Jy).

Single-dish observations of the N$_2$H$^+$\,(1-0 transition were obtained with MOPRA\footnote{MOPRA is operated by the Australian Telescope National Facility (ATNF).} in a survey by \citet{Tackenberg2014}. The observations have a beam FWHM of 35\,$\farcs$5 and velocity resolution of 0.11\,$\kms$.

We combined the PdBI and MOPRA data within the GILDAS software package, using the {\tt UVSHORT} task in natural weighting mode. The synthesised beam of the combined dataset is 7.21'' $\times$ 2.97'' with a position angle of 6 degrees east of north. At the final velocity resolution of 0.2 km s$^{-1}$, the 3$\sigma$ rms is 15\,mJy\,beam$^{-1}$.

We reduced the data with the CLASS module of the GILDAS software package.  We fit a low-order polynomial to the baseline of each spectra, and the spectra were then summed over the relevant regions. We implemented the {\tt hfs} method to account for the hyperfine structure of this transition \citep[e.g.][]{Caselli1995}.

\subsection{IRAM 30-meter: C$^{18}$O~~2 $\rightarrow$ 1}
\label{ss:co}

We observed the region in the C$^{18}$O(2-1) transition at 219.560354\,GHz with the IRAM 30-meter telescope at Pico Veleta in Granada, Spain in January 2012 (project 172-11) in good conditions with less than 1\,mm of precipitable water vapour. Pointing was checked with observations of 1757-240, and focus was performed on Mercury. 

At this frequency, the telescope beam is 11$\farcs$8. Mapping was done in on-the-fly mode with position-switching, using a fixed OFF position $\alpha_{2000}$ = 18:10:16.6, $\delta_{2000}$= -19:00:59.9. The HERA receiver has 1\,GHz bandwidth on the FFTS backend with which we achieved a velocity resolution of 0.25\,$\kms$ and 1-$\sigma$ rms of 200\,mK in the $T_A^*$ scale. The data were reduced using the CLASS module of the GILDAS software package. Spectral fits were made assuming a Gaussian line profile after removing a first order baseline. These data are presented in full, along with maps of the atomic and ionised carbon line transitions, in \citet{Beuther2014}.

\subsection{Continuum studies}
\label{ss:cont}

We incorporate multiple continuum measurements from previous studies to constrain the evolutionary stage of the cores in the region. Single-dish measurement at 850\,$\mu$m with SCUBA shows that this region (box in Figure 1a) contains $\sim$700\,$\msun$ of gas mass \citep{A&ASpecialIssue-Henning}.  The {\em Herschel} \citep{A&ASpecialIssue-Herschel} guaranteed time key project entitled the Earliest Phases of Star Formation \citep[EPoS:][]{A&ASpecialIssue-Henning, Ragan2012b} obtained maps at all PACS \citep{A&ASpecialIssue-PACS} and SPIRE \citep{A&ASpecialIssue-SPIRE} wavelengths of G11. The PACS data (see Figure~\ref{fig:finder}) in particular are sensitive to compact, protostellar cores embedded in the dense gas reservoir. We use these far-infrared data to infer the evolutionary stage of the various emission peaks depending on the characteristics of the spectral energy distribution (SED). Region 1 (Figure~\ref{fig:finder}, panel b) is infrared-dark through 160\,$\mu$m, which signifies a very high column density and/or a lack of an internal heating source. Region 2 is dark at 24\,$\mu$m, but bright at all PACS bands, which places it at a more advanced evolutionary stage than region 1. Region 3 is bright in all infrared bands, indicating that it is the most evolved source. The SEDs ($\lambda \geq$ 70\,$\mu$m) of the main cores in regions 2 and 3 were modelled with blackbody functions in \citet{A&ASpecialIssue-Henning} resulting in temperatures of the 'cold' core component (containing the bulk of the mass) of 19 and 21\,K, respectively. With {\em Herschel} detections only at 100 and 160\,$\mu$m, we placed an upper limit of 16\,K on the temperature in region 1.

The SEDs of cores were revisited in \citet{Ragan2013} with the addition of high-resolution (7$\farcs$8) SABOCA data at 350\,$\mu$m. Regions 1 and 3 were detected in the SABOCA maps, while region 2 lacked sufficient signal-to-noise. The cores are of intermediate mass, between 7 and 17\,$\msun$, and bolometric temperatures \citep[$T_\mathrm{bol}$;][]{MyersLadd1993} progress from $<$13\,K for region 1, 19\,K and 37\,K for regions 2 and 3, respectively. The updated properties derived from SED fits are summarised in Table~\ref{tab:her_sab}. For the calculations of masses and column densities, we use the cold component temperatures of 13, 19 and 15\,K for regions 1, 2 and 3, respectively, from \citet{Ragan2013}.

Simultaneously with the PdBI line observations, we obtained 3.2\,mm continuum maps using 3.6\,GHz effective bandwidth of the WideX correlator (Figure~\ref{fig:finder}b). The beam was 6$\farcs$88 $\times$ 2$\farcs$74, and we achieved a 3\,$\sigma$ sensitivity of 0.17\,mJy\,beam$^{-1}$ and detected regions 2 ($\alpha$,$\delta$=18$^h$10$^m$16.3$s$, $-$19$^{\circ}$:24$^{'}$:20.4$^{''}$) and 3 ($\alpha$,$\delta$=18$^h$10$^m$19.2$s$, $-$19$^{\circ}$:24$^{'}$:22.9$^{''}$) only. These regions have effective diameters of 7$''$ and 11$''$, respectively, in the continuum image.  

\begin{table*}
\begin{center}
\caption{Continuum properties of {\em Herschel}-identified regions. \label{tab:her_sab}}
\begin{tabular}{ccc|cc|cc|ccrl}
\hline \hline
Region & RA (J2000) & Dec (J2000) & $S_{3.2\mathrm{mm}}^\mathrm{tot}$ & $M_{3.2\mathrm{mm}}$  &   $S_{350{\mu}\mathrm{m}}^\mathrm{tot}$ & $M_{350{\mu}\mathrm{m}}$ & $L_\mathrm{bol}$ & $T_\mathrm{bol}$ & $N(\rm{H}_2)^a$~~ & Description \\
Number & [$^{h}$:$^{m}$:$^{s}$] & [$^{\circ}$:$^{'}$:$^{''}$] & [mJy] & [$\msun$] & [Jy] & [$\msun$] &  [$\lsun$] & [K] & [10$^{22}$\,cm$^{-2}$]  & \\
\hline
1 & 18:10:17.9 & -19:24:37 & $< 0.17$ & $<$~1 & 2.0 & 14$^a$ & $< 5^a$ &  $< 13^a$ & $>$ 8.4$\pm$1.1   & MIR-dark \\  
2 & 18:10:16.2 & -19:24:19 & 0.4 & 3$\pm$1 & $\dots$ & 7$^b$ & 9$^b$ & 19$^b$ & ~~~~$\dots$~~~~ & 24\,$\mu$m-dark\\           
3 & 18:10:19.4 & -19:24:24 & 2.4 & 17$\pm$3 & 2.1 & 17$^a$ &  27$^a$  &  37$^a$ & 5.6$\pm$0.8  & 24\,$\mu$m-bright \\ 
\hline
\end{tabular}
\end{center}

\tablefoottext{a}{From \cite{Ragan2013}, assuming cold component temperature.}\\
\tablefoottext{b}{From \cite{A&ASpecialIssue-Henning}.}

\end{table*}

\section{Results: Dense cores}
\label{s:dense}

\subsection{Continuum emission}
\label{ss:contresults}

Figure~\ref{fig:finder} provides an overview of the G11 filament. Panel $a$ shows the large-scale filament at 70\,$\mu$m from our {\em Herschel}/PACS observations with SPIRE 350\,$\mu$m contours overlaid \citep{Ragan2012b}. In panel $b$, we show the PdBI 3.2\,mm continuum emission over the {\em Spitzer}/MIPS 24\,$\mu$m image. The 3.2\,mm continuum is detected toward 70\,$\mu$m-bright Regions 2 and 3, totalling 0.4 and 2.4\,mJy, respectively. Following the standard assumptions\footnote{Assuming an extrapolated dust opacity ($\beta$ = 1.7) from column 5 ($n$ = 10$^6$\,cm$^{-3}$, thin ice mantles) of $\kappa_\mathrm{3mm}$ = 0.23\,cm$^2$\,g$^{-1}$ \citet{ossenkopf_henning}, gas-to-dust mass ratio of 100, cold-component temperatures from \citet{Ragan2013}.}, we estimate masses from these continuum measurements, arriving at 3 and 17\,$\msun$ for Regions 2 and 3, respectively. The error in the mass assumes the temperatures are good within 1\,K \citep{Ragan2012b} and does not account for uncertainty in the dust opacity. The latter can influence the mass estimate by up to a factor of two. We do not detect 3.2\,mm continuum toward 70\,$\mu$m-dark Region 1 (3$\sigma$ sensitivity of 0.17 mJy beam$^{-1}$, corresponding to 1\,$\msun$). We note that if the extinction distance of 4.7\,kpc were adopted, then the mass values would be about a factor of two higher.

For comparison, in $c$, we show the 350\,$\mu$m SABOCA continuum emission \citep{Ragan2013}, which has similar angular resolution of 7.8$''$ plotted over {\em Herschel}/PACS 100\,$\mu$m data. Regions 1 and 3 are detected in the SABOCA map, and although 350\,$\mu$m emission is seen near region 2, it is not a significant 3-$\sigma$ detection (rms = 0.27\,Jy beam$^{-1}$). The continuum properties are summarised in Table~\ref{tab:her_sab}. Overall, the continuum selects the three bright peaks that had been identified with {\em Herschel}, while the N$_2$H$^+$\,(1-0) emission (see Section~\ref{ss:n2hp}) is more extended.

\subsection{N$_2$H$^+$\,1$\rightarrow$0 emission}
\label{ss:n2hpresults}

We show the integrated N$_2$H$^+$(1-0) emission in panels $d$ (PdBI-only) and $e$ (PdBI+MOPRA) of Figure~\ref{fig:finder}. In both, the N$_2$H$^+$ exhibits significantly more structure than the continuum. Infrared-identified regions 1 through 3 (see Section~\ref{ss:cont}) lie at the peaks of extended structures in N$_2$H$^+$. We identify additional peaks in N$_2$H$^+$ integrated emission (e.g. denoted 1a, 1b, etc., in order of decreasing peak intensity) which were above our threshold of 0.75\,Jy beam$^{-1}$\,$\kms$ in the combined dataset and list their offsets and spectral properties derived using the {\tt hfs} method in the GILDAS package CLASS in Table~\ref{tab:lineparams}. 

We list the core diameters in Table~\ref{tab:lineparams} integrating out to where the N$_2$H$^+$ emission reaches the level of 0.75\,Jy\,beam$^{-1}$\,$\kms$. The main regions (Table~\ref{tab:her_sab}) have a projected separation of 0.45\,pc (between Region 1 and each of the other two regions). The N$_2$H$^+$ sub-cores (Table~\ref{tab:lineparams}) exhibit a mean nearest neighbour separation of 0.18\,pc (0.25\,pc if we had assumed distance of 4.7\,kpc). We note that the Jeans fragmentation length (0.2\,pc assuming 10$^{4}$\,cm$^{-3}$, 12\,K media) is only marginally resolved by our PdBI observations (0.12\,pc $\times$ 0.05\,pc). 

The cores have a mean N$_2$H$^+$ column density of 7 $\times$ 10$^{12}$\,cm$^{-2}$ over their areas and vary by less than a factor of two. Likewise, the [N$_2$H$^+$/H$_2$] abundance is roughly 5 $\times$ 10$^{-10}$ (using $N$(H$_2$) from Table~\ref{tab:her_sab}) in rough agreement with results from both nearby cores \citep[e.g.][]{Caselli2002, taf04,Tobin2013a, Miettinen2013b} and more massive IRDC regions \citep[e.g.][]{ragan_msxsurv, BeutherHenning2009, Vasyunina2011, Sanhueza2013, Gerner2014, Henshaw2014}. 

\begin{table*}
\begin{center}
\caption{Cores and line parameters from PdBI-only and PdBI+MOPRA N$_2$H$^+$ (1-0) datasets. \label{tab:lineparams}}
\begin{tabular}{lccccccccccc}
\hline \hline
& & Angular & Linear & \multicolumn{2}{c}{PdBI only} & & \multicolumn{3}{c}{PdBI+MOPRA} \\
\cline{5-6} 
\cline{8-10} 
Core & Offset$^a$ & Size$^b$ & Size$^c$ & $\mathrm{v}_\mathrm{lsr}$ & fwhm & & $\mathrm{v}_\mathrm{lsr}$ & fwhm & $\tau$ &  & $M_\mathrm{vir}^d$ \\
     & [$''$,$''$] & [$''$] & [pc] & [km s$^{-1}$] & [km s$^{-1}$] & & [km s$^{-1}$] & [km s$^{-1}$] & &  & [$\msun$]\\
\hline
1a & -0.75, -12.00 & 10 & 0.17 & 29.7$\pm$0.2  & 0.53$\pm$0.67 & & 29.7$\pm$0.03 & 0.73$\pm$0.05 & 7.1 &&  6.5 \\ 
1a$'$   &          &    &      & 32.2$\pm$0.2 &  1.19$\pm$0.67 & & 32.0$\pm$0.08 & 1.56$\pm$0.02 & 0.1 && 29.5 \\ 
1b & -8.25, -22.50 & 5  & 0.08 & 31.6$\pm$0.02 & 0.58$\pm$0.04 & & 31.3$\pm$0.04 & 0.99$\pm$0.08 & 1.8 &&  6.0 \\ 
\hline
2a & -23.25, 3.75  & 6  & 0.10 & 30.3$\pm$0.03 & 0.89$\pm$0.05 & & 30.1$\pm$0.04 & 1.05$\pm$0.06 & 11.2 &&  8.0 \\ 
2b & -27.75, 7.50  & 4  & 0.07 & $\dots$       & $\dots$       & & 29.8$\pm$0.08 & 1.48$\pm$0.24 & 1.7 && 10.6 \\ 
\hline
3a & 23.25, 3.00   & 10 & 0.17 & 30.4$\pm$0.01 & 1.45$\pm$0.06 & & 30.1$\pm$0.02 & 1.89$\pm$0.06 & 3.8 && 43.2 \\ 
3b & 9.75,  3.00   & 9  & 0.15 & 30.0$\pm$0.03 & 0.61$\pm$0.04 & & 29.7$\pm$0.03 & 1.28$\pm$0.09 & 7.9 && 17.9 \\ 
3c & 33.75, 5.25   & 5  & 0.08 & 30.3$\pm$0.02 & 0.52$\pm$0.03 & & 30.1$\pm$0.03 & 1.80$\pm$0.11 & 1.9 && 19.7 \\ 
3d & 42.00, 2.25   & 5  & 0.08 & 31.8$\pm$0.03 & 2.08$\pm$0.20 & & 30.2$\pm$0.06 & 2.70$\pm$0.17 & 2.8 && 44.1 \\ 

\hline
\end{tabular}
\end{center}

\tablefoottext{a}{Offsets from the map centre position 18$^h$10$^m$17.9$^s$ -19\degr24$'$24\farcs9.}

\tablefoottext{b}{Average angular diameter of core in the merged dataset.}

\tablefoottext{c}{ Linear diameter of core in the merged dataset assuming a distance of 3.41\,kpc.}

\tablefoottext{d}{The virial mass is computed assuming a $\rho \propto R^{-1.8}$ density profile \citep{Shirley2002} and using the size and linewidth from the merged dataset. }

\end{table*}

\subsection{Velocity structure}

The spectrum at each peak position was fit using the PdBI-only data as well as the combined data cube. The parameters are listed in Table~\ref{tab:lineparams}. While the inclusion of the single-dish MOPRA data does not alter the centroid velocities beyond the measurement error (except for core 3d), the linewidths in the combined data cube are always larger than the PdBI-only data cube by at least 20\% and up to factor of 3, due to the larger beam of the single-dish observations.

\begin{figure*}
\centerline{
\includegraphics[width=2.5in]{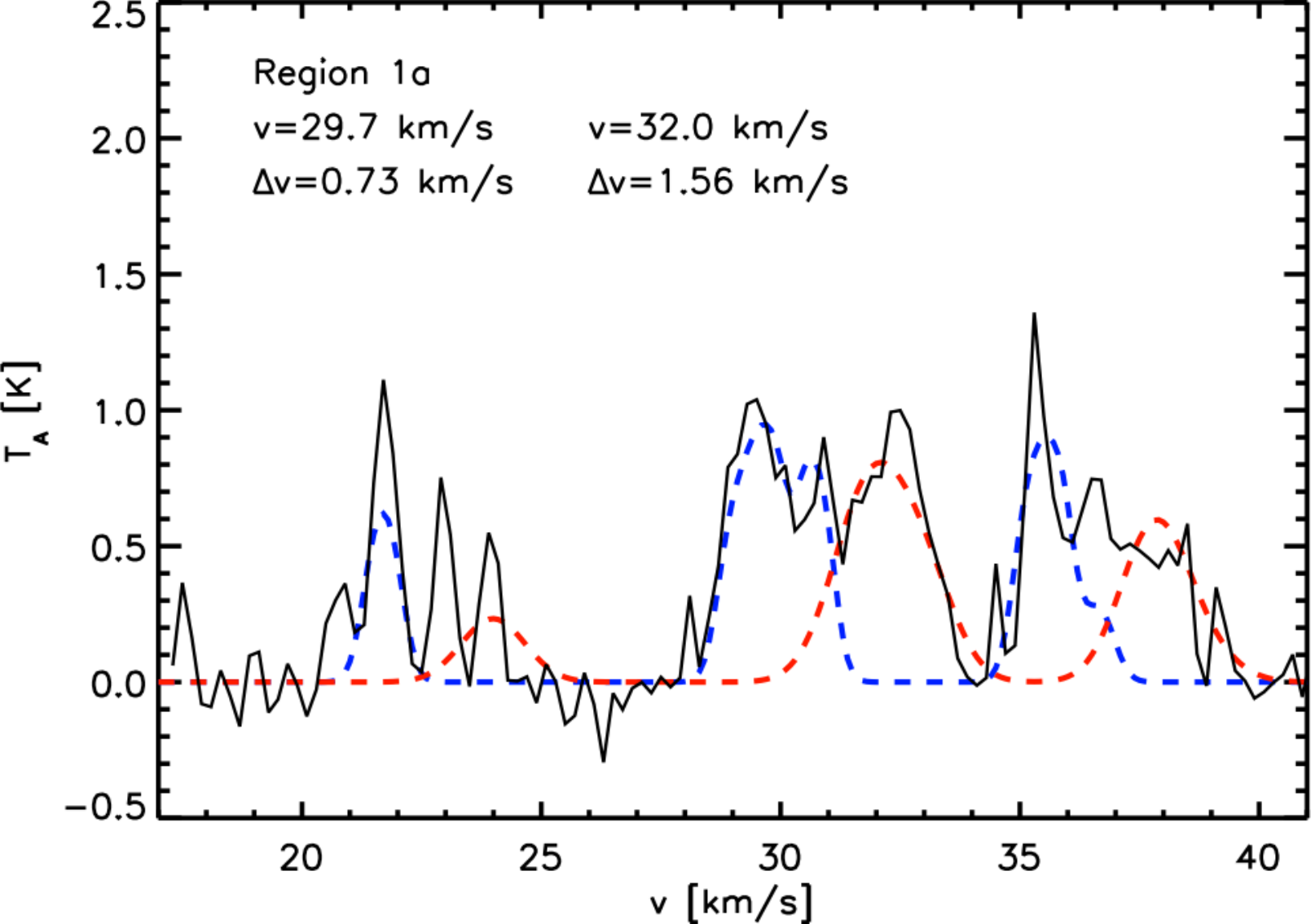}
\hspace{0.5in}
\includegraphics[width=2.5in]{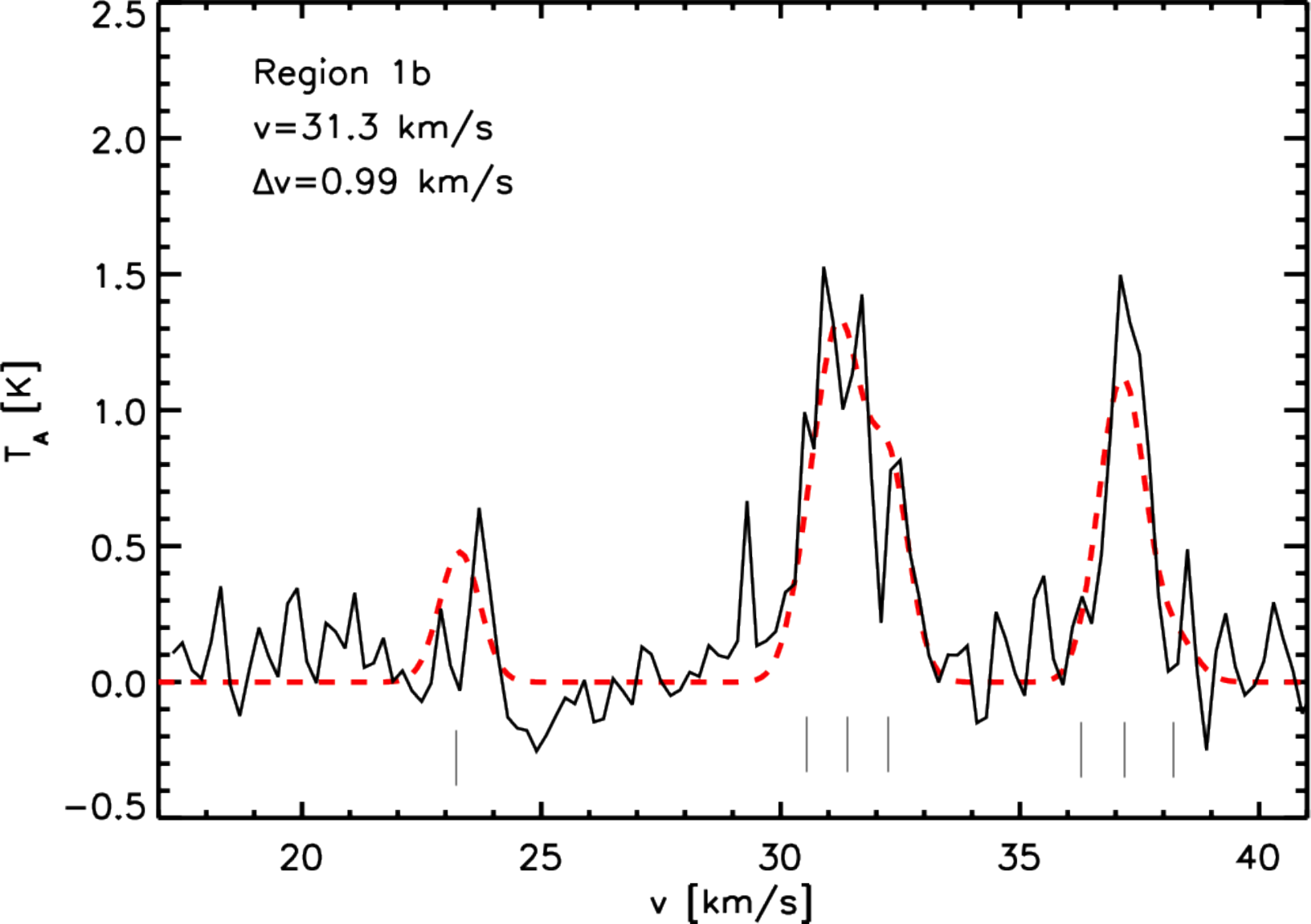}
}
\centerline{
\includegraphics[width=2.5in]{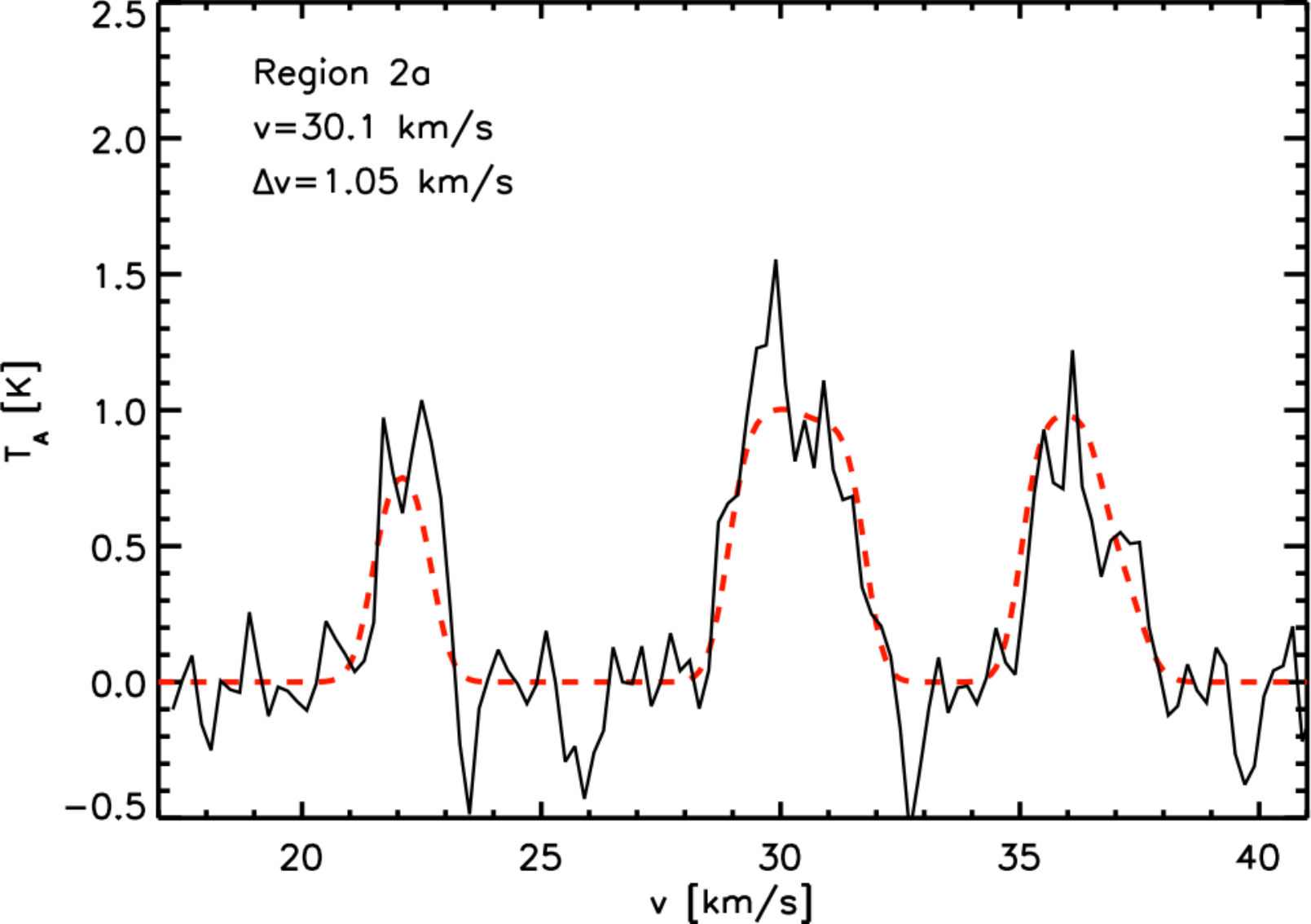}
\hspace{0.5in}
\includegraphics[width=2.5in]{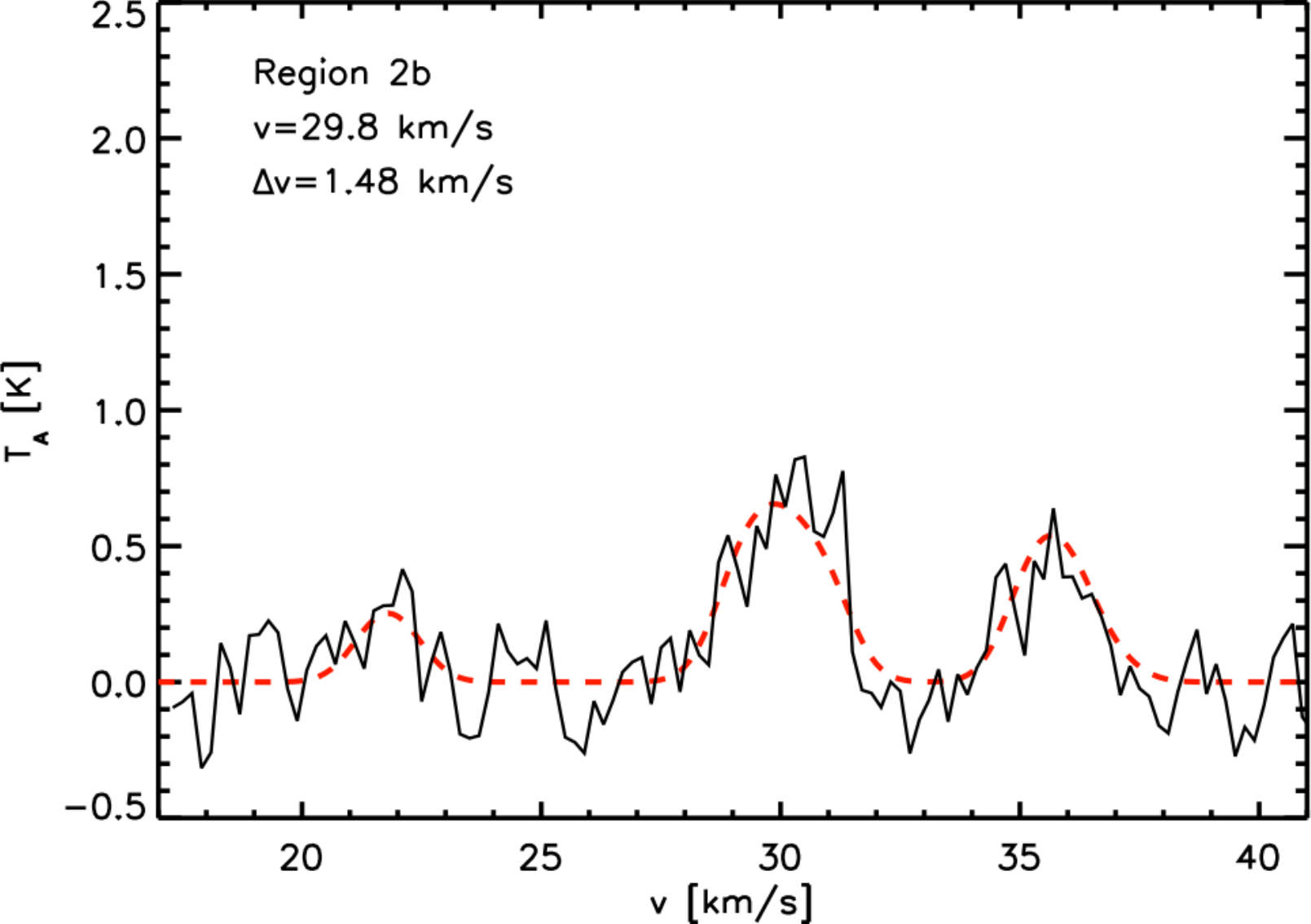}
}
\centerline{
\includegraphics[width=2.5in]{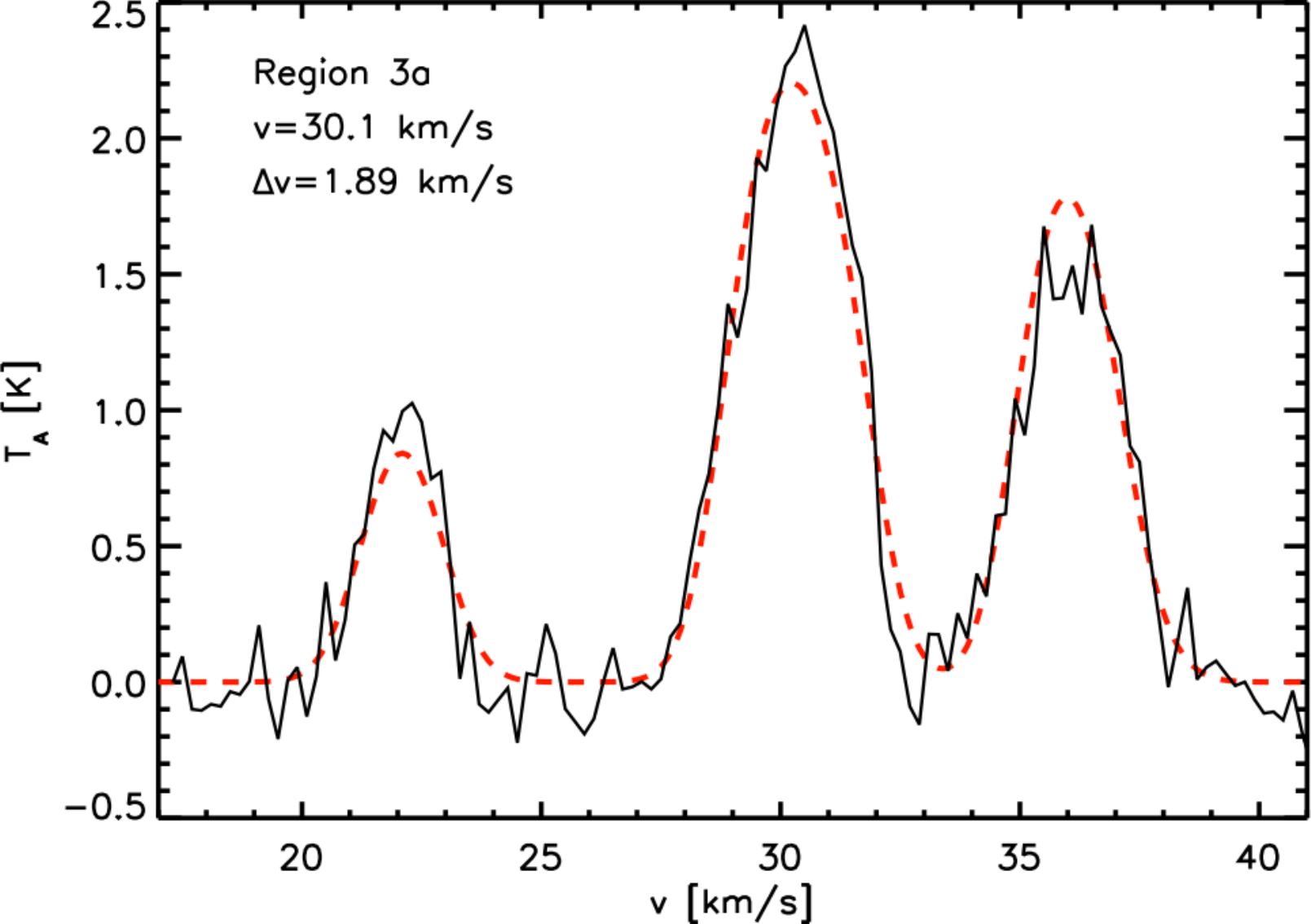}
\hspace{0.5in}
\includegraphics[width=2.5in]{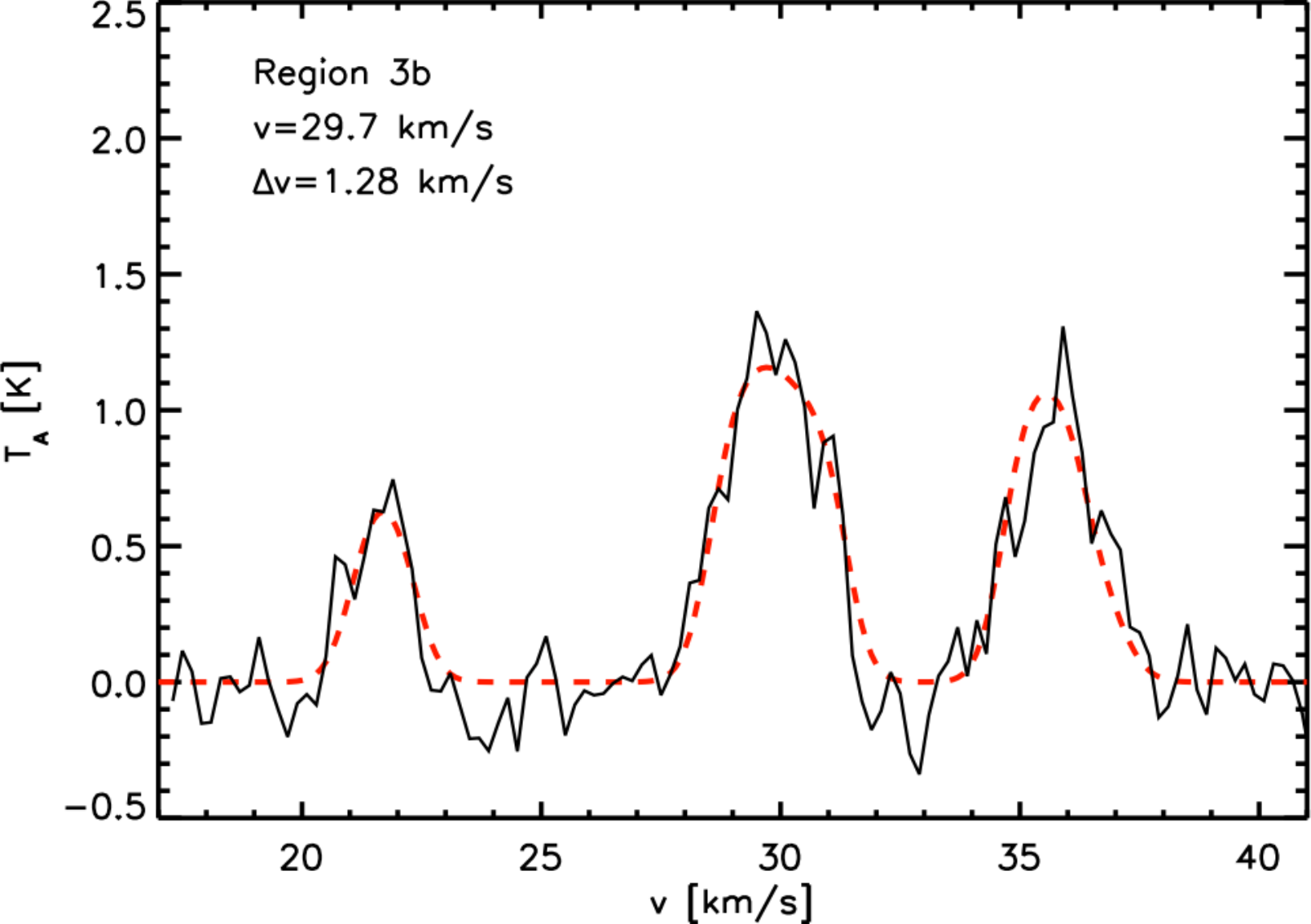}
}
\centerline{
\includegraphics[width=2.5in]{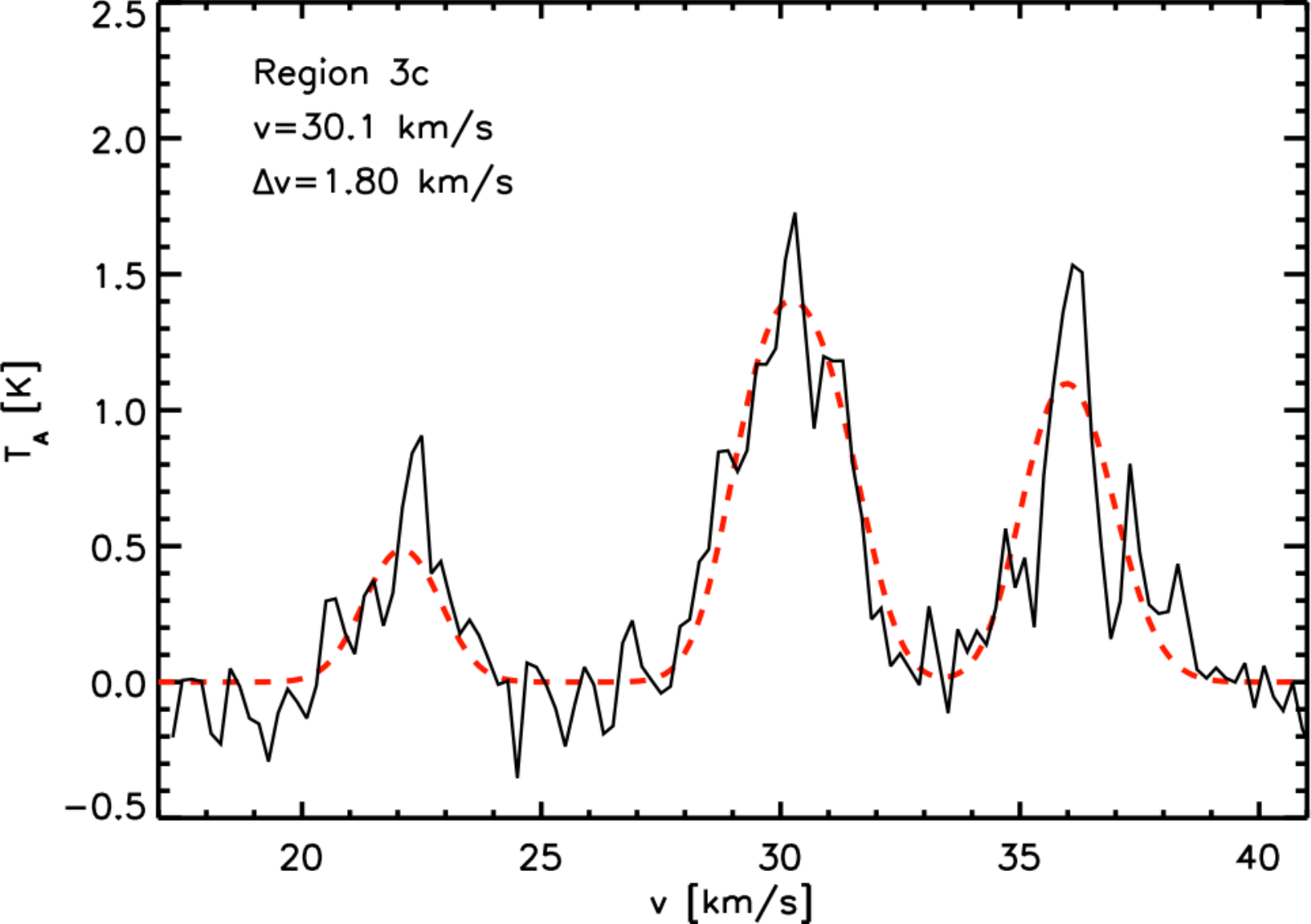}
\hspace{0.5in}
\includegraphics[width=2.5in]{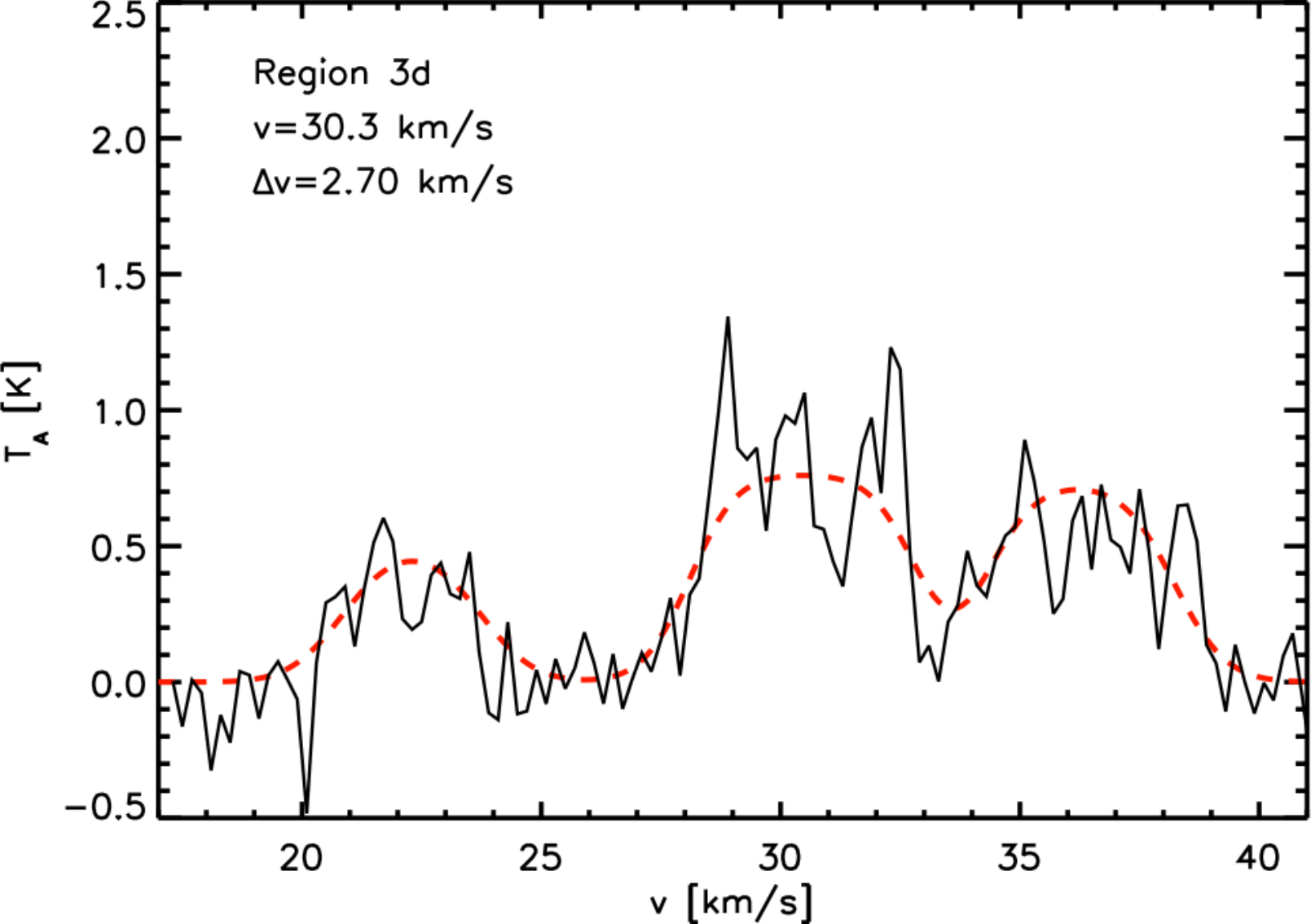}
}
\caption{\label{f:n2hp_spec_fits}N$_2$H$^+$\,1$\rightarrow$0 spectra from the merged PdBI+MOPRA dataset (black lines) and the best fit parameters (red dashed lines, see Table~\ref{tab:lineparams}). The positions seven hyperfine components \citep{Caselli1995} are shown in the top-right panel.}
\end{figure*}

\begin{figure*}
 \includegraphics[angle=0,width=\linewidth]{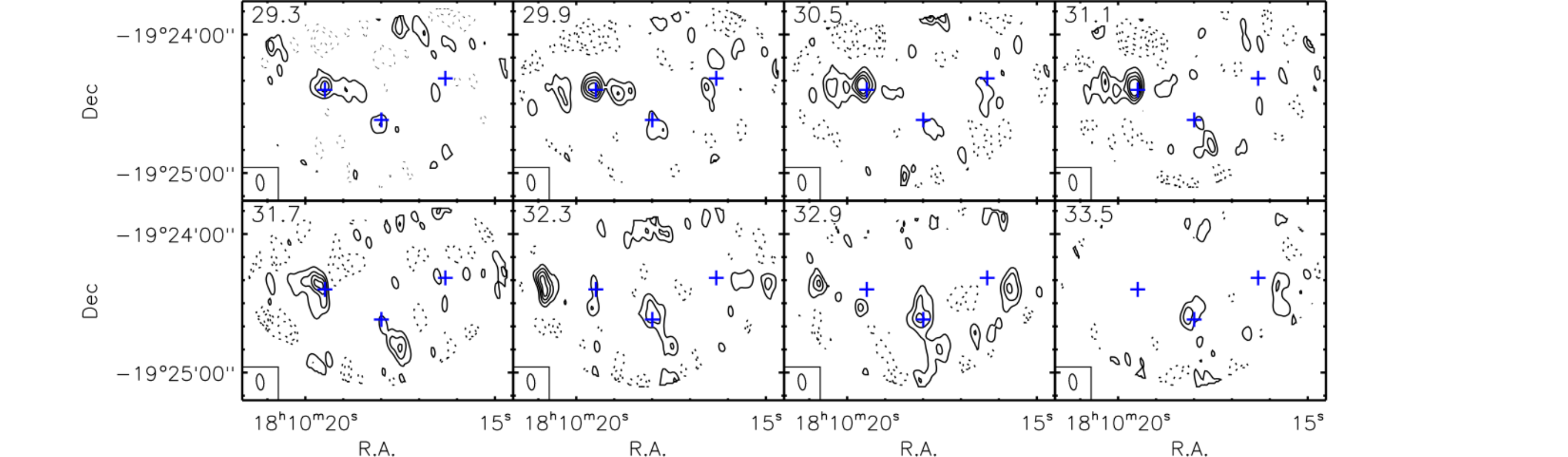}
\caption{\label{f:chanmap} Channel map of merged N$_2$H$^+$ emission. Contours are plotted at -50 (dotted), 50, 100, 150, and 200\,mJy beam$^{-1}$. The velocity channel is indicated in the upper-left of each panel in units of km s$^{-1}$, and the blue plus signs indicate the centres of the main regions. The synthesised beam in shown in the lower-left corner of each panel.}
\end{figure*}

In Figure~\ref{f:n2hp_spec_fits}, we show the N$_2$H$^+$ spectra extracted at each peak position from the combined (PdBI+MOPRA) data cube. The line intensities, velocities and linewidths are given in each panel and in Table~\ref{tab:lineparams}. The strongest emission originates from the 24\,$\mu$m-bright source 3a. Most peaks are well-fit with a single velocity component. A notable exception is 1a, which clearly requires two velocity components for a satisfactory fit. The stronger component is centred near the common velocity of the other cores, 29.7\,km s$^{-1}$, but there is a second component centred at 32.0\,km s$^{-1}$. This is also seen in Figure~\ref{f:chanmap}, which is the channel map of the N$_2$H$^+$ (1-0) emission from the PdBI+MOPRA merged data cube. The steps in velocity are 0.6 km s$^{-1}$. Interestingly, Core 1b exhibits a single velocity component, but nearer to the offset velocity, $\sim$2\,km s$^{-1}$ different from the rest of the cores.

Core 2b is not significantly detected in the PdBI-only maps, which we interpret to mean it is not strongly peaked core.  Core 3d is centred at 31.8\,km s$^{-1}$ in the PdBI-only map but in the combined map it is centred at 30.3\,km s$^{-1}$. Its spectrum (see Figure~\ref{f:n2hp_spec_fits}) shows not only its noisy nature but also a hint of two velocity components.  Its position at the edge of our PdBI map probably is to blame for the low signal-to-noise, because of which a two-component model was not successful.

\section{Results: Core environment traced by C$^{18}$O}
\label{s:environs} 

We use the emission from the 2$\rightarrow$1 transition of C$^{18}$O as a probe of the environment or ``envelope'' within which the dense cores discussed above reside \citep[cf.][]{wmb04,Kirk2007}. The cores themselves are of sufficiently high density that CO is preferentially frozen out onto dust grains, eliminating the main destroyer of N$_2$H$^+$ \citep[e.g.][]{Bergin2002}. In this region of G11, our 11$\farcs$8 resolution C$^{18}$O\,(2-1) maps show no significant small-scale peaks in common with the continuum or N$_2$H$^+$, but rather a more extended morphology, similar to the SPIRE 350\,$\mu$m emission (see Figure~\ref{fig:finder}a). As such, we treat the C$^{18}$O as a tracer of the envelope. 

\begin{figure*}
\centerline{
\includegraphics[width=2.4in]{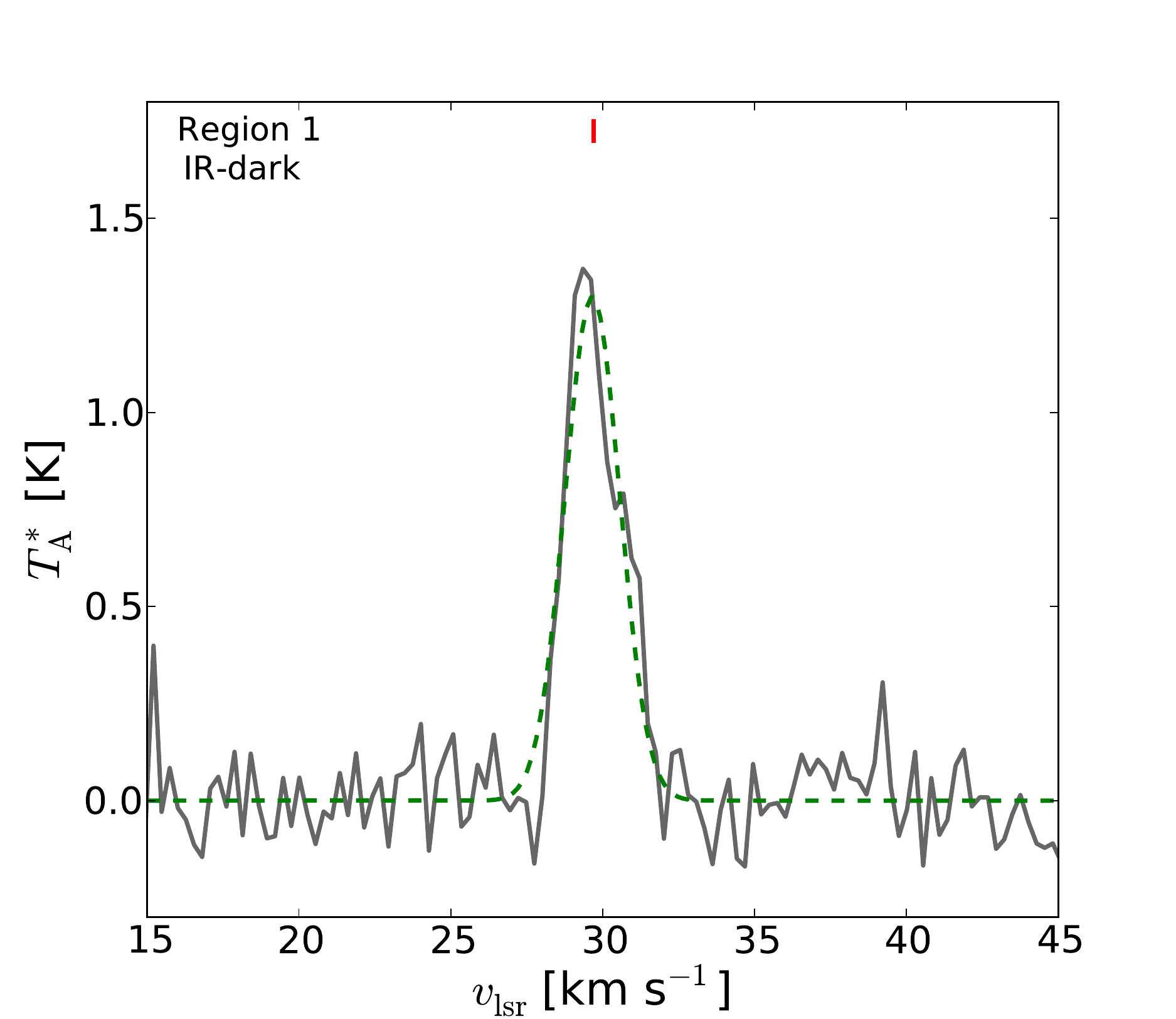}
\includegraphics[width=2.4in]{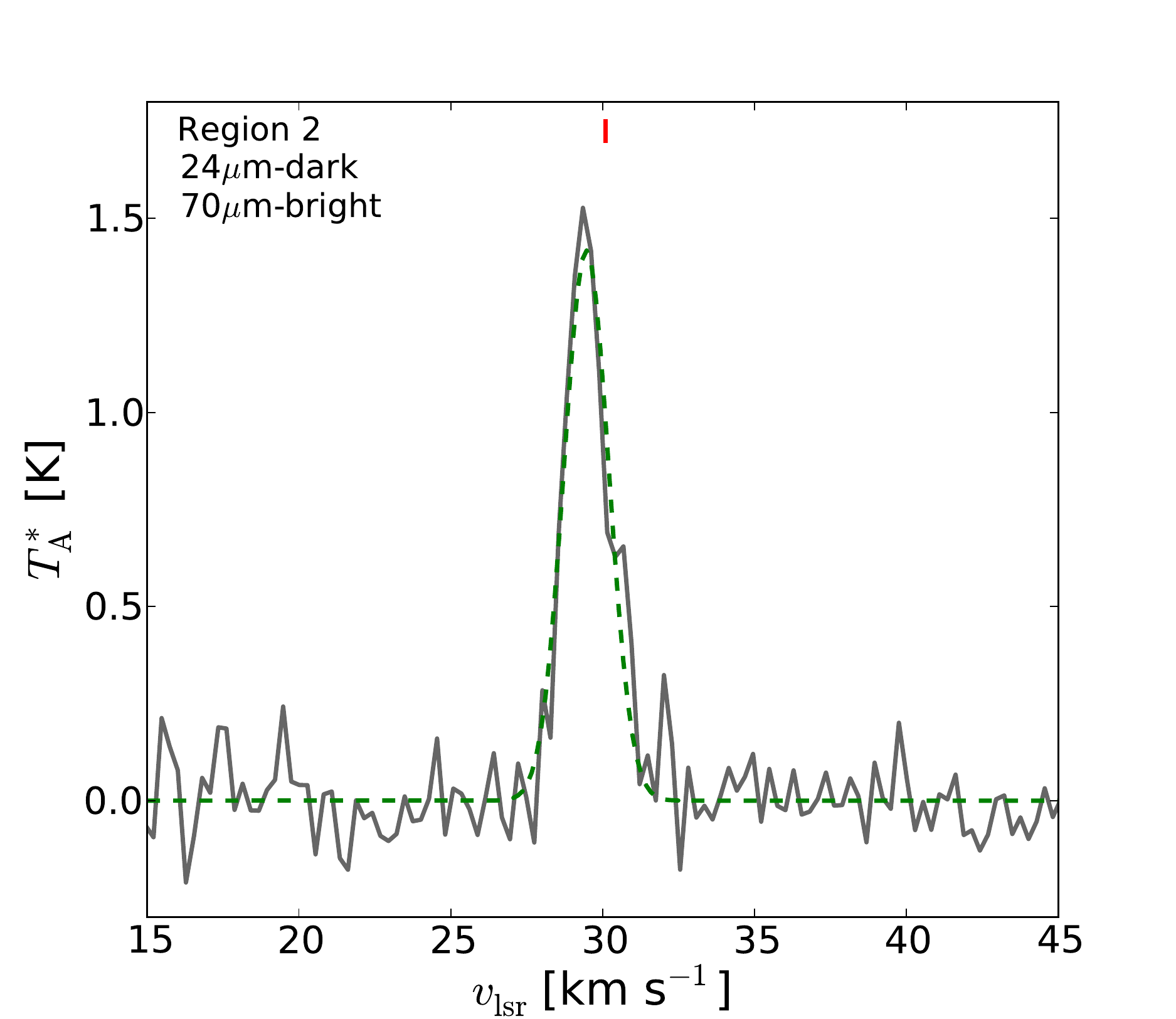}
\includegraphics[width=2.4in]{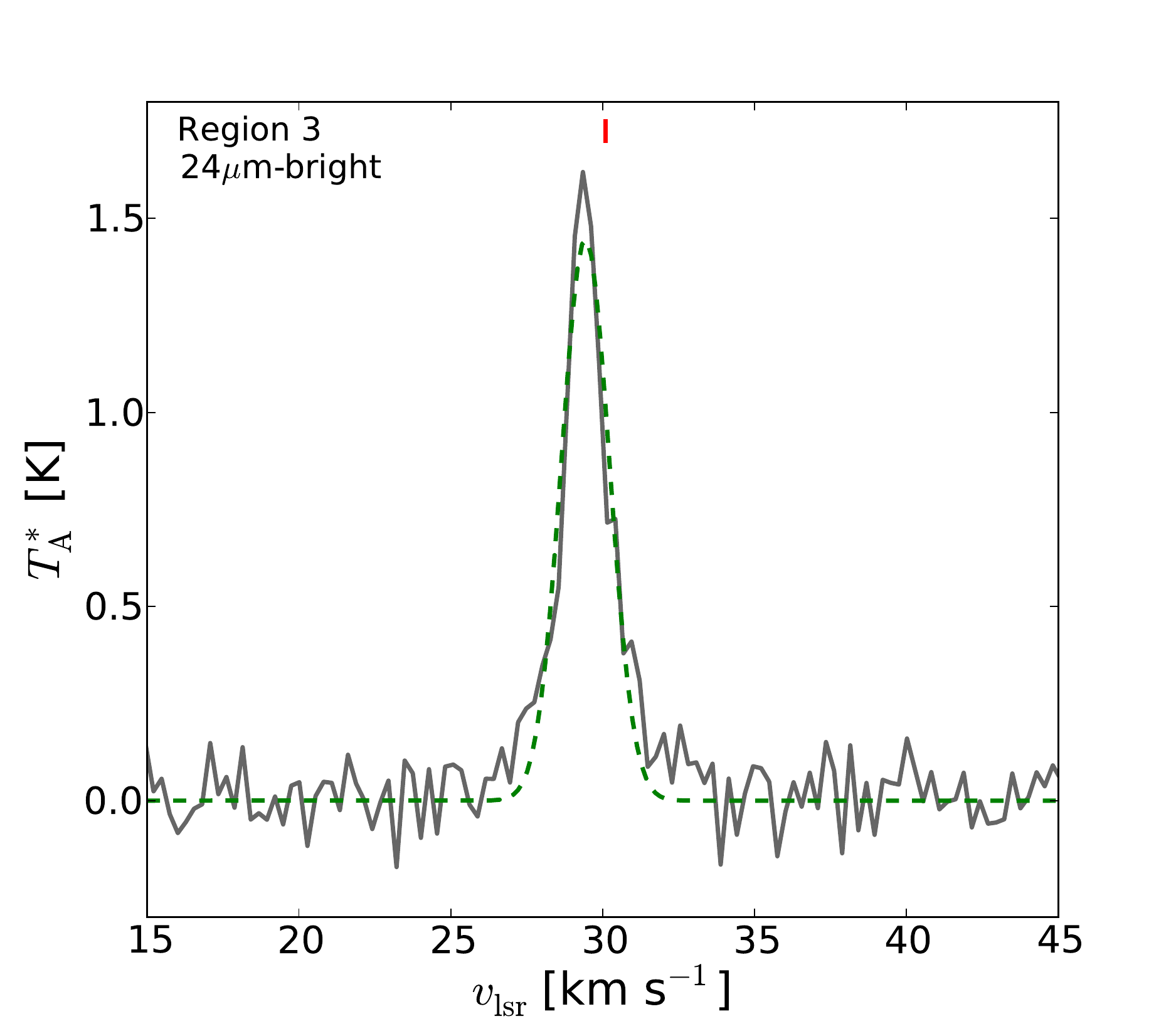}
}
\caption{\label{fig:cospectra} Solid lines show spectra of C$^{18}$O\,(2-1) (green) from the IRAM 30-meter telescope averaged over the 12$''$ regions (see Table~\ref{tab:her_sab}), and the dashed green lines show the best single component Gaussian fits to the data. The fit parameters are summarised in Table~\ref{tab:C18Olineparams}. The red vertical line indicates the velocity centroid from the fits to the PdBI+MOPRA N$_2$H$^+$\,(1-0) data.}
\end{figure*}

By comparing the C$^{18}$O emission to core properties probed by N$_2$H$^+$ discussed above, we will measure the relationship between the cores and their environment. We show the C$^{18}$O\,(2-1) spectra extracted at the positions of the three regions (averaging over 11$\farcs$8 diameter circle) in Figure~\ref{fig:cospectra}. A single Gaussian component can satisfactorily fit the C$^{18}$O\,(2-1) line, the parameters of which are given in Table~\ref{tab:C18Olineparams}. Region 3 exhibits a hint of wing emission, which may be consistent with an outflow from an internal source. We experimented with two-component fits in an attempt to address the slight blue asymmetry, but this did not improve the fit quality \citep[cf.][]{JimenezSerra2014}.

The velocity centroids of the C$^{18}$O\,(2-1) are consistently between 29.4 and 29.7\,km\,s$^{-1}$, consistent within uncertainties with the centroids in the N$_2$H$^+$ spectra. The C$^{18}$O\,(2-1) line is broadest in Region 1, where two velocity components are detected in N$_2$H$^+$\,(1-0), but no counterpart to the 32\,$\kms$ component seen in core 1a is seen in C$^{18}$O \footnote{We note that new observations of [CI] of this region show broadened line profile ($\Delta$v = 4.5\,km\,s$^{-1}$), corresponding to a broadening of C$^{18}$O\,(2-1) line, 25$''$ (0.4\,pc) south of Region 1 \citep{Beuther2014}.}. In summary, the C$^{18}$O\,(2-1) emission is fairly uniform over this region, with a mean $T_\mathrm{mb}$\,=\,0.93\,K, v$_\mathrm{lsr}$\,=\,29.52\,km\,s$^{-1}$ and $\Delta$v\,=\,1.90\,km\,s$^{-1}$ (fwhm).

\begin{table}
\begin{center}
\caption{Regions and the line parameters from IRAM 30 m observations of C$^{18}$O\,2$\rightarrow$1. \label{tab:C18Olineparams}}
\begin{tabular}{ccccc}
\hline \hline
Region  &  $T_\mathrm{A}^*$ & v$_\mathrm{lsr}$ & fwhm  &  \\
  & [K] & [km s$^{-1}$] & [km s$^{-1}$]  & \\
\hline
1 & 1.30$\pm$0.17 & 29.66$\pm$0.04 & 2.12$\pm$0.09 & \\ 
2 & 1.42$\pm$0.13 & 29.46$\pm$0.03 & 1.73$\pm$0.09 & \\ 
3 & 1.45$\pm$0.13 & 29.43$\pm$0.03 & 1.85$\pm$0.08 & \\ 
\hline
\end{tabular}
\end{center}
\end{table}

%
%
\section{Discussion}
\label{s:disc}

\subsection{Virial analysis}
\label{ss:virial}
The virial mass ($M_\mathrm{vir}$) provides an estimate of how gravitationally bound a core is: a core which exceeds its virial mass by a factor of 2 or more is considered self-gravitating. We assume an underlying density profile of $\rho \propto R^{-1.8}$, where $R$ is the core radius, for the cores based on the median value found by \cite{Shirley2002} in their study of Class 0 protostars. Following \citet{MacLaren1988}, we calculate the virial mass for cores with the assumed density profile, the expression for which is $M_\mathrm{vir}$ = 147~$R$~($\Delta$v)$^2$, where $R$ is the core radius in parsecs evaluated in the combined N$_2$H$^+$ map, and $\Delta$v is the N$_2$H$^+$ linewidth (fwhm) in km s$^{-1}$. The derived values, which are listed in Table~\ref{tab:lineparams}, range from 6\,$\msun$ in Region 1 to 44\,$\msun$ in Region 3. These values would be elevated to 8 and 60\,$\msun$ if G11 is instead assumed to be at the extinction distance, 4.7\,kpc.

We compare the submillimetre mass ($M_{350{\mu}\mathrm{m}}$, Table~\ref{tab:her_sab}) in the main regions to the virial masses computed at the main (``a'') peaks in the N$_2$H$^+$ combined PdBI+MOPRA maps. We find a virial parameter ($\alpha_\mathrm{vir}$ = $M_\mathrm{vir}/M_{350{\mu}\mathrm{m}}$) of 1.1 and 2.5 (1.5 and 3.5 for 4.7\,kpc distance) for Regions 2 and 3, respectively, values which are close to the ``critical'' boundary of $\sim$2 below which objects are considered unstable to collapse. As there are two velocity components in Region 1, we estimate virial masses for each: 6.5\,$\msun$ (29.7\,km s$^{-1}$ component) and 29.5\,$\msun$ (32.0\,$\kms$ component). Because the mass is summed in that line of sight, it is not possible to say how much mass corresponds to each velocity component. Either way, $M_{350{\mu}m}$ of Region 1 is within a factor of $\sim$2 of the virial mass. These $\alpha$ values are typical of what is observed in high-mass regions throughout the literature \citep{Kauffmann2013b}.    

As was pointed out by \cite{Ballesteros2006} in the context of a turbulent ISM, we caution the reader that considering the cores in isolation is a flawed approach in determining core stability. Indeed the cores may represent the sites of local collapse favoured in such an environment. If we consider the boundedness of the entire region of radius $R\sim$0.8\,pc (approximating as a sphere), which exhibits an average linewidth of 1.9\,km\,s$^{-1}$ ($\sigma$ = 0.81\,km\,s$^{-1}$) in C$^{18}$O, and contains $\sim$700\,$\msun$ from dust continuum emission \citep{A&ASpecialIssue-Henning}, the virial mass, given in this case by $M_\mathrm{vir}$ = 126\,$R$\,($\Delta$v)$^2$ (assuming $\rho \propto R^{-2}$) is $\sim$360\,$\msun$, resulting in a virial parameter $\sim$0.5. From this, we would conclude that this region is likely undergoing collapse from large scales, similar to global trends seen by \citet{Ragan2012}. However, we emphasise again that this simplistic analysis neglects any contribution from magnetic fields, uncertainties in the dust emissivity and other effects, any of which could contribute factors of a few to this analysis. 

\subsection{Fragmentation scale}
\label{ss:fragmentation}
We find the average nearest-neighbour separation between our sample of N$_2$H$^+$ cores (see Table~\ref{tab:lineparams}) to be 0.18\,pc. The cores have effective radii ($r_\mathrm{eff}$ = $\sqrt{\mathrm{area}/\pi~}$) ranging between 0.04 and 0.08\,pc. Our observations are not sensitive to the sub-structure of cores, but rather probe how the large-scale IRDC filament sub-divides from clumps to cores \citep[cf.][]{Kainulainen2013, Wang2014}. As G11, is a well-known filamentary IRDC, it is tempting to attribute the nature of its fragmentation to the so-called ``sausage instability'', which arises from the collapse of an infinite isothermal cylinder \citep[e.g.][]{ChandrasekharFermi1953, Ostriker1964}.  \citet{Kainulainen2013} showed that on large length scales ($l >$ 0.5\,pc), a collapsing isothermal cylinder reasonably predicts the structure separations, but on smaller scales such as those probed in this work, thermal Jeans instability analysis appears a better match.  

We first consider the region as a whole. The temperature in this region is $\sim$12\,K \citep{Pillai_ammonia}, the total mass is 700\,$\msun$ and the mean density is 10$^4$\,cm$^{-3}$ \citep{A&ASpecialIssue-Henning}. The thermal Jeans fragmentation length\footnote{In the context of non-uniform and supersonic environment of IRDCs, the Jeans length should be understood as an upper limit for equilibrium \citep{StahlerPalla}. The supersonic conditions in IRDCs, which may be driven by accretion from large scales \citep{Klessen2010, Heitsch2013}, violate the  assumptions inherent in basic Jeans analysis \citep{Goodman1998}. Other processes, such as turbulence, rotation or support from magnetic fields, could be influencing the observed structure as well.} applied at this scale is then $\sim$0.2\,pc, which agrees with the observed structure. The Jeans mass is 4\,$\msun$.
The cores themselves have mean densities of 1-2 $\times$ 10$^5$\,cm$^{-3}$ and slightly higher temperatures (see Table~\ref{tab:her_sab}, taking 20\,K as a representative temperature), which predicts a fragment size of 0.09\,pc, which agrees with the observed sizes, and mass of $\sim$2\,$\msun$. We see that thermal Jeans fragmentation satisfactorily predicts the sizes and separations of the cores, similar to what has been observed in other young regions \citep[e.g.][]{Teixeira2006,Takahashi2013,Beuther2013}. That $M > M_\mathrm{Jeans}$ indicates that, in the absence of other means of support, we expect the cores to fragment further beyond the resolution of our observations.

\subsection{Core evolution signatures}
\label{ss:core_evolution}

Our continuum studies of this region have provided us with a means with which to assess the evolutionary stage of the cores, i.e. regions 1, 2 and 3. Table~\ref{tab:her_sab} lists the bolometric temperature \citep[$T_\mathrm{bol}$,][]{MyersLadd1993}, which was determined by integrating the spectral energy distribution (SED) over all available data and deriving the characteristic temperature. This method has proven especially useful in analysing new protostellar populations discovered in {\em Herschel} data \citep[e.g.][]{Stutz2013,Ragan2012b}. Simply, as a core evolves, its $T_\mathrm{bol}$ increases. Due to the small sample size under discussion here, we merely use $T_\mathrm{bol}$ as a proxy for relative evolutionary stage and do not speculate further whether the trends seen in the line emission hold universally. Strictly speaking, all cores fall below the fiducial $T_\mathrm{bol}$ = 70\,K threshold proposed by\citet{Andre1993}, qualifying all sources as Class 0 cores, but as was demonstrated in \citet{Ragan2013} and \citet{Stutz2013}, $T_\mathrm{bol}$ faithfully follows the evolution within this phase, such that more evolved cores have higher values of $T_\mathrm{bol}$.

We consider the kinematic properties of the cores in the context of their evolutionary stage as probed by the infrared characteristics. If we assume that continuum regions 1, 2 and 3 (Table~\ref{tab:her_sab}) correspond most closely with emission from N$_2$H$^+$ cores 1a, 2a and 3a (Table~\ref{tab:lineparams}), then the linewidth of N$_2$H$^+$ clearly increases with advancing evolution, i.e. increasing $T_\mathrm{bol}$. The C$^{18}$O linewidth is larger in region 3 than in region 2, but broadest in starless region 1. The latter may be due to unresolved multiple components (though not at the same separation in velocity as N$_2$H$^+$ core 1a and 1a$'$, see Section~\ref{ss:motions}), but as it stands no correlation is apparent in C$^{18}$O. Inspection of $^{13}$CO\,(2-1) spectra at the same positions (not shown) exhibit identical linewidths over the entire region of $\sim$4\,$\kms$.

%
\subsection{IRDC core kinematics}
\label{ss:motions}

There is general agreement between observations and cluster formation theories that star-forming cores are found at over-densities in turbulent molecular clouds, however the dynamical aspect of this condition is still debated. Growing observational evidence \citep[e.g.][]{Csengeri2011a,Ragan2012,Peretto2013} has demonstrated that global infall and collapse are prevalent in the early stages of (high-mass) star formation, challenging the assumptions of quasi-static initial conditions \citep[e.g.][]{mckee_tan03}. However, if we are to adopt a more dynamical framework of the initial conditions of cluster formation, observational constraints of the kinematics on all scales are still lacking. In the following, we address several aspects of the dynamical environment.

\subsubsection{Multiple velocity components in N$_2$H$^+$}
In massive IRDCs, N$_2$H$^+$(1-0) emission is excited in extended regions, and when observed with a single dish telescope the derived linewidth is typically supersonic\citep[cf.][]{Tackenberg2014}. However, when we restrict our view to scales below 0.1\,pc (core scales), we often find that due to the exclusion of unrelated or extended material, the line widths are much narrower, enabling us to isolate multiple velocity components, the most striking example of which presents itself in core 1a (and 1a$'$). The separation of the line centroids is about 2.5\,km\,s$^{-1}$: much larger than separations observed in low-mass filaments \citep{HacarTafalla2011, HacarTafalla2013}, slightly larger to what was seen (at the same spectral resolution) in 4 out of 17 cores observed in intermediate mass IRDC\,19175 by \citet{BeutherHenning2009} and in IRDC filaments in G035.39-00.33 \citep{Henshaw2014}, but similar to velocity centroid separations observed in high-mass IRDCs \citep[e.g.][]{Csengeri2011b,Beuther2013,Peretto2013}. 

The ubiquity of multiple velocity components reported in the literature undeniably points to the importance of the dynamical picture in the early stages of cluster formation, on both large and small scales. In most of the studies mentioned above, the two velocity peaks in N$_2$H$^+$ are seen over parsec spatial scales, so it is possibly worth distinguishing these cases from instances in which the two components of dense gas tracers are restricted to individual cores such as \cite{BeutherHenning2009} and this work. Our target region in G11 does not exhibit multiple components separated by $>$\,1\,km\,s$^{-1}$ on large filamentary scales, but rather seems to be confined to core 1a and 1a'. Using simulations of isolated clouds, \citet{Smith2013} assert that such a signature in optically thin species, such as N$_2$H$^+$, can be simply explained by multiple density peaks along our sightline. However, our C$^{18}$O\,(2-1) data shows that no counterpart to the second velocity component detected in N$_2$H$^+$, which would be expected if structure on the line of sight were the only cause. Another possibility for the origin of the second component in N$_2$H$^+$ is shocked emission arising from dense material gravitationally in-falling onto the young core 1a, possibly connected to the global collapse of the whole region. Further multi-tracer observations at $\sim$0.1\,pc resolution are required to determine the prevalence of this signature in dense, young objects.

\subsubsection{Core motions}
Core-to-core velocity dispersions in clouds may be a useful diagnostic in ascertaining whether turbulence is driven or decaying \citep{Offner2008a}. In simulations with turbulent driving, prestellar cores displayed a higher core-to-core velocity dispersion than protostellar cores, and the reverse was true in simulations with decaying turbulence. Observational evidence for both scenarios exists: while Perseus \citep{Kirk2007} agrees with driven turbulence, Ophiuchus \citep{Andre2007} and NGC\,2068 \citep{WalkerSmith2013} agree better with the decaying turbulence scenario. 

Our target region in the G11 IRDC offers an interesting case of intermediate and low-mass cores embedded in a large mass reservoir. Unfortunately, due to the limited region we were able to map, our sample size is too small to draw any strong conclusions. Nevertheless, assuming that this region in G11 is oriented roughly in the plane of the sky, the overall standard deviation of centroid velocities among the cores (see Table~\ref{tab:lineparams}) is $\sigma$ = 0.87\,$\kms$ (PdBI-only values) and 0.74\,$\kms$ (PdBI+MOPRA values). We find no significant variation between the line centroids of the two prototellar cores (1a and 2a, $\sigma$ = 0.05\,km\,s$^{-1}$), and the core-to-core velocity dispersions for (apparently) prestellar/starless cores is higher, $\sigma$ = 0.97\,km\,s$^{-1}$ (including core 1a$'$, 0.83\,$\kms$ for PdBI+MOPRA). Since $\sigma_\mathrm{prestellar} > \sigma_\mathrm{protostellar}$, this region follows the trend seen for driven turbulence simulations, although this should be repeated with a much larger statistical sample of cores. Additionally, massive IRDC filaments are highly turbulent environments which were not the focus of the \cite{Offner2008a} study, so we emphasise that this is not firm evidence against the decaying turbulence scenario.

We also compare the core motions with respect to the large scale gas reservoir traced by C$^{18}$O. While the C$^{18}$O is fairly uniform over this region, we note that the linewidth is most enhanced at the position of the two N$_2$H$^+$ velocity components (region 1). We do not, however, detect a corresponding velocity component to N$_2$H$^+$ core 1a with velocity $\sim$32\,$\kms$; if a second C$^{18}$O unresolved velocity component exists there, it is at a velocity closer to the systemic $\sim$30\,$\kms$. Does this second N$_2$H$^+$ component indicate that the emitting core (1a$'$) is somehow decoupled from the larger mass reservoir? Or does it have a different origin altogether? 

With the exception of core 1a$'$, the N$_2$H$^+$ core velocity centroids agree remarkably well with the velocity of C$^{18}$O emission (their common reservoir). In the dynamical models posed by \citet{Bonnell2001a,Bonnell2001b},\citet{Bonnell2006}, and \citet{Ayliffe2007}, this coupling between cores and their reservoir indicates that the region is in an early stage at which cores can accrete effectively. This is similar to what has been observed in relatively low-mass environments \citep{wmb04, Kirk2007, Kirk2010, WalkerSmith2013}, which generally find little or no offset between the velocity centroids of the high-density gas tracer and the tracer of the less dense mass reservoir, typically differing by less than the local sound speed.


\section{Conclusions}
\label{s:conclusion}

We present new observations of a quiescent region in the IRDC G011.11-0.12. This region contains three intermediate mass starless/protostellar cores which have previously been characterised by {\em Herschel} and SABOCA. We have mapped the region in N$_2$H$^+$\,(1-0) with the Plateau de Bure Interferometer along with the 3.2\,mm continuum at 7.21$''$ $\times$ 2.97$''$ (0.12 $\times$0.05\,pc) resolution. We mapped the C$^{18}$O\,(2-1) transition with the IRAM 30-meter telescope at 11$\farcs$8 ($\sim$0.2\,pc) resolution. We summarise our results as follows:

\begin{itemize}

\item We find the continuum emission traces the three primary, {\em Herschel}-identified cores. The continuum peaks also mark the maxima in integrated N$_2$H$^+$ emission. The N$_2$H$^+$ emission appears as a chain of cores with radii from 0.04 to 0.08\,pc. The mean nearest neighbour separation is 0.18\,pc. Both the core separations and their sizes are well-predicted by isothermal Jeans fragmentation, similar to other detailed studies at similar physical resolutions. 

\item We find that the main continuum-identified regions have masses between factors 1.1 to 2.5 times less than the virial mass, placing them at the boundary between sub- and super-critical, but similar to the low-values of the virial parameter ($\alpha_\mathrm{vir}$) cores observed in high-mass regions. However, we emphasise that the regions are located in a vast mass reservoir, which invalidates the basic assumptions of the virial theorem. Turning our attention to the stability of the whole region traced by single-dish observations, it is supercritical ($\alpha_\mathrm{vir}$ $\sim$ 0.5) and is thus likely undergoing large-scale gravitational collapse. We note that our analysis does not account for the contribution of magnetic fields or geometrical effects, which are needed for a more comprehensive understanding of the region's stability.

\item Using the bolometric temperature ($T_\mathrm{bol}$) as a proxy for relative evolutionary stage of the cores, we find the N$_2$H$^+$ linewidth increases with evolutionary stage, while the C$^{18}$O shows a more modest increase, and $^{13}$CO shows none at all. This suggests that the feedback from embedded protostars is quite localised and/or at an early stage before prominent molecular outflows are evident.

\item The starless region 1, with $M_{350\mu m}$ = 14\,$\msun$ and $T_\mathrm{bol}$\,$<$\,13\,K, exhibits two clear velocity components in N$_2$H$^+$ in both the PdBI-only map and combined PdBI+MOPRA map, separated by 2.5 and 2.3\,$\kms$, respectively. Emission at the velocity of the second component is seen to extend south to another quiescent core (1b). Such two-component signatures do not appear in the N$_2$H$^+$ spectra of more evolved regions 2 and 3 and also do not have a counterpart in the C$^{18}$O\,(2-1) spectrum. This observation amplifies the importance of not only the complex structure of young regions, but also the complex dynamics, such as global gravitational collapse or shocks, in understanding the initial phases of cluster formation. 

\item The core velocities (from N$_2$H$^+$) match the characteristic velocity of the C$^{18}$O, the probe of the common mass reservoir. The core-to-core dispersion is of the same order as the dispersion of the C$^{18}$O\,(2-1) line. With the exception of core 1a$'$, the cores appear to be coupled to their common mass reservoir such that accretion can effectively proceed.

\end{itemize}

\begin{acknowledgements}

We thank Jochen Tackenberg for kindly providing reduced MOPRA data, Anike Schmiedeke for supplying the {\em Herschel} EPoS team with reduced data, and Paul Clark for useful discussions. We thank the referee and editor for helpful comments which helped to improve the paper. SER acknowledges support from the Deutsche Forschungsgemeinschaft priority program 1573 (``Physics of the Interstellar Medium''). This research has made use of NASA Astrophysics Data System. This research made use of APLpy, an open-source plotting package for Python hosted at {\tt http://aplpy.github.com}.
      
\end{acknowledgements}


\end{document}